\documentclass[prb,showpacs]{revtex4}
\usepackage{amsmath}
\usepackage{amssymb}
\usepackage{graphicx}
\usepackage{bm}

\input{tcilatex}

\begin{document}

\title{Trion dynamics in coupled double quantum wells.
Electron density effects.}
\author{P. Aceituno}
\email{paceitun@ull.es}
\author{A. Hern{\'{a}}ndez-Cabrera}
\affiliation{Departamento de F\'{\i}sica B\'{a}sica, Universidad de La
Laguna, La Laguna 38206-Tenerife, Spain.}
\date{\today }

\begin{abstract}
We have studied the coherent dynamics of injected electrons when they are
either free or bounded both in excitons and in trions (charged excitons). We
have considered a remotely doped asymmetric double quantum well where an
excess of free electrons and the direct created excitons generate trions. We
have used the matrix density formalism to analyze the electron dynamics for
different concentration of the three species. Calculations show a
significant modification of the free electron inter-sublevel oscillations
cWe have studied the coherent dynamics of injected electrons when they are
aused by electrons bound in excitons and trions. Based on the present
calculations we propose a method to detect trions through the emitted
electromagnetic radiation or the current density.
\end{abstract}

\maketitle


\section{Introduction}

Coherent oscillations (quantum beats) of the electronic charge in double
quantum wells (DQW's) have been widely studied both experimentally and
theoretically \cite{1,2,3,4,5,6,7,8}. The most usual method to achieve
charge generation in these structures is the photoexcitation by an
ultrashort laser pulse, which causes the same concentration of photoexcited
electrons and holes and so, the corresponding excitons. Another possibility
consists on injecting electrons and holes by doping two regions close to the
structure \cite{9,10}. Injection offers the advantage of preventing the
interaction between the electromagnetic field associated to the excitation
and the excited electrons. Here excitons are directly created by
hole-assisted electron resonant tunneling (see Fig. 1). Another
characteristic of this type of excitons (the so-called direct-created
excitons) is that they have spatial coherence, with an in-plane momentum $%
\mathbf{k}_{exc\parallel }\sim \mathbf{0}$, and can radiate in the direction
perpendicular to the DQW. Thus, direct-created excitons are good candidates
for their application in vertical planar microcavities. Moreover, injection
has two additional advantages: the control of the electron and hole
densities by means of the $n$ (donor) or $p$ (acceptor) impurity
concentration plus an external applied electric field \cite{10b}, and the
avoidance of exciton thermalization. This thermalization is an inevitable
consequence of non-tuned laser excitation.

\begin{figure}
\begin{center}
\includegraphics{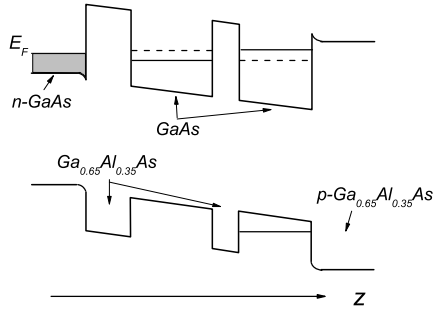}
\end{center}
\par
\addvspace{-1 cm}
\caption{Scheme of the asymmetric double quantum well structure under free
electron resonant condition.}
\label{fig.1}
\end{figure}

Recently, trions (charged excitons) have been observed in photoexcited doped
semiconductor quantum wells. The doping is necessary to get an excess of
electrons. These electrons together with the excitons lead to the generation
of the negatively charged excitons\cite{10b,11,11b}. These bound complexes
of three particles have a binding energy large enough to make them
observable. The experimental technique most widely used to study trions is
time resolved photoluminescence\cite{12,13}. Processes as generation and
recombination of excitons and trions have been recently analyzed through
this technique\cite{14}.

In this work we propose a new method to generate trions in coupled DQW
through the direct-created excitons and the necessary free electron excess.
In the usual experimental conditions, donor concentration is greater than
acceptor concentration ($\sim 10^{18}$ and $\sim 10^{16}$ cm$^{-3}$,
respectively). The process of trion generation can be considered in three
steps: first, if the Fermi level of the $n$-type material (see Fig. 1)
resonates with the electronic level in the left-hand wide quantum well
(LQW), electrons tunnel to LQW forming the cross or spatially indirect
(interwell) excitons with holes diffused into the narrow right-hand well
(RQW). Second, if the resonance condition for electrons in the excitonic
state between both wells is satisfied, electrons in the excitonic state
tunnel to the RQW forming spatially direct (intrawell) excitons. Third, if
there is an excess of electrons with respect to holes, free electrons will
coexist together with direct-created excitons. Some of these free electrons
can be bound to excitons leading to trions. It should be noted that, in
coupled quantum wells, the strong dipole-dipole repulsion between excitons
prevents the formation of biexcitons. Thus, we will neglect the
contributions from these neutral species \cite{15}.

Under resonance conditions the charge density in coupled DQWs oscillates
between both wells leading to an oscillating dipole moment and the
corresponding Terahertz radiation emission. If we consider the holes
confined in the well where they were initially injected (due to their large
effective masses), the electronic charge will play the main role in the
coherent process. Because we will deal with three different types of
electron binding (free electron, excitonic electron and trionic electron)
they will occupy different energy levels. Moreover, the resonance conditions
will be slightly different for each case and thus, their oscillatory
behavior will be different as well.

The aim of this paper is the theoretical analysis of the electron in trion
dynamics in presence of direct-created excitons and free electrons. We will
carry out this in the framework of the matrix density formalism, taking into
account the different generation, recombination and annihilation rates for
electrons with different type of binding. The dynamics of coexisting
excitons and free electrons was studied elsewhere \cite{10}. The work is
organized as follows: in Sec. I we present the exciton and trion
Hamiltonians and the corresponding Schr\"{o}dinger equations, as well as the
dynamics of the electronic density evolution. In Sec. II we include
numerical results for the charge density dynamics and for the current
density. We also include some comments regarding the results and the used
approximations.

\section{Formalism}

Initially, a group of electrons tunnels into the LQW. The charge,
dynamically trapped within the wells, produces a reaction field that
modifies the structure profile and then, the time evolution of the system.
Thus, we need to calculate wave functions for the free-electron, the
excitonic electron and the trionic electron, and the corresponding
electronic levels, by solving self-consistently the one-particle Schr\"{o}dinger equation, together with the Poisson equation for each species.

Several techniques have been used to solve the Schr\"{o}dinger equation for
neutral excitons and charged excitons (trions).  We can cite the \textit{commutation technique}, developed by Combescot et al.\cite{16}. Although it
is a very attractive formalism, it is almost inapplicable from the numerical
point of view. Another technique, based on the deformed correlated Gaussian
wave functions, is the \textit{Stochastic variational method} used by Riva
et al.\cite{17}. Whittaker and Shields\cite{18} used a set of \textit{Landau-level basis }to study the in-plane $\left( xy\right) $ motion. These
two methods are specially indicated when an external magnetic field is
applied. In presence of longitudinal electric fields and in absence of
magnetic fields Dacal and Brum \cite{19} have developed a technique based on
the \textit{configuration interaction method (CIM) }\cite{20}. \ Because of
its feasibility we will use this last technique with some implementations.

The \textit{CIM} considers \ that the longitudinal ($z$) and transversal $%
\left( xy\right) $ parts of the wave functions are separable. Thus, we first
factorize wave functions of excitonic and trionic electrons into the
corresponding in-plane part and confinement (growth) direction part. We
can't neglect the in-plane transversal motion part of the Hamiltonian
because both parts of the split Hamiltonian are connected by the Coulomb
potential. Such a circumstance affects wave functions of the different types
of electrons. In order to reach the resonance for the different electronic
states we have included in calculations an external electric field applied
in the growth direction. The confinement part of the Hamiltonian can be
solved by means of the Airy functions within the effective mass approach for
electrons and holes. To solve the in-plane part of the one particle
Hamiltonian we will follow the same steps as in Ref. \cite{19}. First we
will use polar coordinates in terms of the center of mass (CM) and relative
to the hole coordinates. To say,
\begin{eqnarray}
\mathbf{\rho }_{1} &=&\mathbf{\rho }_{e1}-\mathbf{\rho }_{h},  \nonumber \\
\mathbf{\rho }_{2} &=&\mathbf{\rho }_{e2}-\mathbf{\rho }_{h},  \nonumber\\
\mathbf{\rho }_{CM} &=&\frac{\mathbf{\rho }_{e1}+\mathbf{\rho }%
_{e2}+m_{h\perp }\mathbf{\rho }_{h}}{m_{h\perp }+2m_{e}}.
\end{eqnarray}
where $m_{h\perp }$ is the in-plane hole effective mass. We have considered
an isotropic electron effective mass.

\subsection{Exciton and trion Hamiltonians}

The excitonic Hamiltonian can be written as
\begin{equation}
H_{exc}=H(z_{e})+H(z_{h})+H_{\perp }+V_{c},
\end{equation}
where
\begin{equation}
H(z_{e,h})=-\frac{\hbar ^{2}\partial }{2\partial z_{e,h}}\frac{1}{m_{e,hz}}%
\frac{\partial }{\partial z_{e,h}}+V_{DQW}(z_{e,h})+V_{H}(z_{e,h})\pm
\left\vert e\right\vert Fz_{e,h}.
\end{equation}%
Here $m_{hz}$ is the hole effective mass in the growth direction, $
V_{DQW}(z_{e,h})$ is the asymmetric DCW potential, $F$ is the electric field
required to reach the inter-level resonance. $V_{H}(z_{e,h})$ is the Hartree
potential caused by the Coulomb interaction of the electrons with themselves
and with the donors (or holes with acceptors). The self-consistent potential 
$V_{H}(z_{e,h})$ can be obtained by solving the Poisson's equation%
\begin{equation}
\frac{d^{2}V_{H}(z_{e,h})}{dz_{e,h}^{2}}=\frac{4\pi e^{2}}{\epsilon }\left[
N_{D,A}(z_{e,h})-\dsum\limits_{j}n_{j}\left\vert \phi
_{j}(z_{e,h})\right\vert ^{2}\right] ,
\end{equation}
where $N_{D,A}$ is the doping profile; thus, $\int
N_{D,A}(z_{e})dz_{e,h}=\dsum\limits_{j}n_{je,h}=n_{2De,h}$ is the in-plane
averaged $2D_{e,h}$ density of electrons (donors) or holes (acceptors) and $j$ refers to the occupied subbands.

The in-plane term reads 
\begin{equation}
H_{\perp }=-\frac{\hbar ^{2}}{2\mu }\left[ \frac{1}{\rho }\frac{\partial }{
\partial \rho }\left( \rho \frac{\partial }{\partial \rho }\right) +\frac{1}{
\rho ^{2}}\frac{\partial ^{2}}{\partial \rho ^{2}}\right] ,
\end{equation}
where we have considered invariant the in-plane excitonic $\mu $ reduced
mass. The Coulomb potential is

\begin{equation}
V_{ce,h}(\mathbf{\rho })=\int \int dz_{e}dz_{e^{\prime },h}\phi
_{e}^{2}(z_{e})\phi _{e^{\prime },h}^{2}(z_{e^{\prime },h})\frac{e^{2}}{
\epsilon \sqrt{\rho ^{2}+\left\langle z_{e}-z_{e^{\prime },h}\right\rangle
^{2}}},
\end{equation}
and includes electron-hole ($e-h)$ and electron-electron ($e-e^{\prime }$)
interactions. This Coulomb effective potential is valid for excitons and for
trions, being slightly different for each case due to their different
electron wave functions and expected values of $\left\langle z-z^{\prime
}\right\rangle $. \ We will assume a constant dielectric function $\epsilon $
across the interfaces. For practical reasons we change the notation of the
above Coulomb potential to
\begin{equation}
V_{ij}(\mathbf{\rho })=\frac{e_{i}e_{j}}{\epsilon }\int \int dzdz^{\prime }
\frac{\phi _{i}^{2}(z)\phi _{j}^{2}(z^{\prime })}{\sqrt{\rho
^{2}+\left\langle z-z^{\prime }\right\rangle ^{2}}},
\end{equation}
where $e_{i}=-e$ for $i=u,l$ and $e_{j}=e\ (-e)$ for $j=h\ (u,l)$. Indexes $u,l,h$ refer to electrons in the upper or lower levels and to holes in the
right-hand QW, respectively.

The trionic Hamiltonian can be written in an analogous way to the excitonic
case by using relative coordinates for the transversal motion:
\begin{eqnarray}
H_{tr} &=&\dsum\limits_{i=1,2}\left[ H\left( z_{ei}\right) +H_{\perp i}
\right] +H(z_{h})+\frac{1}{m_{h\perp }}\mathbf{p}_{1}\mathbf{p}_{2}+  \nonumber
\\
&&\frac{e^{2}}{\epsilon }\frac{1}{\sqrt{\left\vert \mathbf{\rho }_{1}-
\mathbf{\rho }_{2}\right\vert ^{2}+(z_{e1}-z_{e2})^{2}}}
\end{eqnarray}
where $\mathbf{p}_{i}$ is the in-plane linear momentum for the $i$th
particle,
\begin{equation}
H(z_{ei,h})=-\frac{\hbar ^{2}}{2}\frac{\partial }{\partial z_{ei,h}}\frac{1}{
m_{ei,hz}}\frac{\partial }{\partial z_{ei,h}}+V_{DQW}(z_{ei,h})\pm
\left\vert e\right\vert Fz_{ei,h},
\end{equation}
and
\begin{equation}
H_{\perp i}=-\frac{\hbar ^{2}}{2\mu }\left[ \frac{1}{\rho _{i}}\frac{
\partial }{\partial \rho _{i}}\left( \rho _{i}\frac{\partial }{\partial \rho
_{i}}\right) +\frac{1}{\rho _{i}^{2}}\frac{\partial ^{2}}{\partial \rho
_{i}^{2}}\right] -\frac{e^{2}}{\epsilon \sqrt{\left( z_{ei}-z_{h}\right)
^{2}+\rho _{i}^{2}}}.
\end{equation}

To solve the above Hamiltonians we will follow the same steps as Ref. \cite%
{19}. Thus, through the \textit{CIM} we construct a non-orthogonal basis set
of Slater determinants to solve the eigenvalue problem. This method is
widely described in Ref. \cite{20}. We build up the exciton trial wave
function as
\begin{equation}
\Psi _{exc}(z_{e},z_{h},\mathbf{\rho )}=\dsum\limits_{ijk}c_{ijk}N_{ijk}\phi
(z_{e})\phi (z_{h})\varphi _{k}^{m}(\mathbf{\rho }),
\end{equation}
where $c_{ijk}$ is a variational parameter, $N_{ijk}$ is the determinant
normalization, indexes $ijk$ refers to one particle $i,$ $j$, $k$,
respectively, $\mathbf{\rho =\rho }_{e}-\mathbf{\rho }_{h}$, $m$ is the
relative particle angular momentum, and $\phi (z_{e,h})$ is the envelope
wave function, and $\varphi _{k}^{m}(\mathbf{\rho )}$ is the one particle
in-plane wave function
\begin{equation}
\varphi _{k}^{m}(\mathbf{\rho )}=N_{km}\rho ^{m}\exp \left[ -\frac{\rho ^{2}
}{\lambda _{k}^{2}}\right] \exp \left( im\theta \right) .
\end{equation}
Here $N_{km}$ is the relative particle normalization factor and $\lambda
_{k} $ is a set of parameters (one for each angular momentum $m$) that we
have taken from Ref.\cite{21}.

The above expressions can be generalized for trions as 
\begin{eqnarray}
\Psi _{tr}(z_{e1},z_{e2},z_{h},\mathbf{\rho }_{1},\mathbf{\rho }_{2})
&=&\dsum\limits_{ijmnpqr}c_{ijmnpqr}N_{ijmnpqr}\phi _{p}(z_{h})  \nonumber \\
&&\times \left[ 
\begin{array}{c}
\phi _{q}(z_{e1})\varphi _{i}^{m}\left( \mathbf{\rho }_{1}\right) \phi
_{r}(z_{e2})\varphi _{j}^{n}\left( \mathbf{\rho }_{2}\right) \\ 
+\phi _{r}(z_{e1})\varphi _{j}^{n}\left( \mathbf{\rho }_{1}\right) \phi
_{q}(z_{e2})\varphi _{i}^{m}\left( \mathbf{\rho }_{2}\right)%
\end{array}
\right] .
\end{eqnarray}
It is important to remark that, in absence of external magnetic fields, the
only bound state for trions is the singlet $m+n=0$. We would like to mention
here that we have included a new contribution: the spatial bending of the
structure due to the doping. Thus, after variationally obtaining the above
wave functions for $V_{H}(z_{e,h})=0$, we need to solve the Poisson
equations (4) and turn again to the Slater Hamiltonians. Fortunately, we
only need to modify one particle envelope functions in the growth direction
by solving self-consistently the $z$-direction part of the corresponding Schr\"{o}dinger equations (3, 9) 
together with the Poisson equations. For
practical purposes, to numerically work the $z$-direction part with the
Transfer Matrix Method is better than using the Airy functions. Variational
coefficients and normalization functions for wave functions (11, 13) will be
slightly different when including the spacial self-consistent potential  $
V_{H}(z_{e,h})$. Based on Ref. \cite{20} we use $\lambda _{k}$ parameters
for the in-plane part of the Hamiltonians corresponding to $s$ and $p^{\pm }$
states$.$ The inclusion of the $d^{\pm }$ states only affects about a 5\%
the position of the excitonic and trionic energy levels but it supposes an
additional computational effort which is not reflected on the final results.
Another key point consists on the identification of the levels because it
will be fundamental in the next section. This information is contained in
the envelope functions $\phi (z_{e,h})$. Thus, both for electrons and holes,
the two deepest levels correspond to the resonant levels of the two coupled
wells. Hereinafter we will distinguish them with the indexes $u$ (upper) and 
$l$ (lower). We have neglected the possible hole states mixing for the
applied electric fields under consideration because we are mainly interested
in the conduction band states.

Concerning to free-electron, the electronic levels ($29.97\ meV$ and $36.61\
meV$) would correspond to the LQW and the RQW, respectively, if they were
decoupled. For the excitonic electron case these two levels are shifted by
the Coulomb interaction by different energy amounts corresponding to the
indirect and direct exciton \cite{10} ( $5.78\ meV$ and $8.60\ meV$,
respectively). We have considered that hole energy levels remain the same
due to the bigger hole effective mass. The case of trions is more
complicated due to two possible additional forms of setting the new electron
with regard to the previous excitonic state. From the energy point of view,
the most favorable cases correspond to the direct exciton bounded to an
electron either in the LQW or in the RQW. That means a shift of the free
electron levels of $6.91\ meV$ and $9.93\ meV$, respectively. We represent
in Fig. 2 electronic levels as function of the external electric field. It
can be seen that the resonances of the three species take place for almost
the same applied electric field. With these values we calculate the level
splitting for coupled wells, $\Delta _{T}$, and the transmission matrix
element, $T$, which equals to $\Delta _{T}/2$ at resonance. The level
splitting energy for decoupled wells is $\Delta =\sqrt{\Delta _{T}^{2}-4T^{2}
}$, for each case. We will use these parameters in the next subsection.

\begin{figure}
\begin{center}
\includegraphics{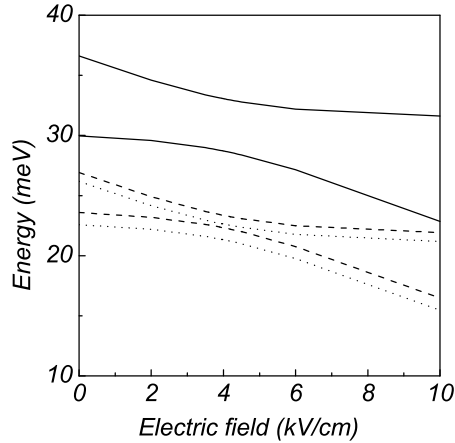}
\end{center}
\par
\addvspace{-1 cm}
\caption{Energy levels versus applied electric fiel. Solid line: free
electrons. Dashed line: excitonic electrons. Dotted line: trionic electrons.}
\label{fig.2}
\end{figure}

\subsection{Time evolution of the electron density}

In this Section we will treat the dynamic behavior of the electronic density
within the Matrix Density formalism in the momentum representation. Although
electrons spend some time in tunneling the left-hand confining barrier, we
assume an ultrafast $\delta (t)$ injection of electrons in the LQW to
simplify calculations. Also, diffusion of holes from the $p$-doped material
to the RQW is considered as a $\delta (t)$ function. We use this
approximation because we are mainly interested in the dynamics of the
electronic density between wells after injection. Based on the previous
premises, we start from the general quantum kinetic equation for the density
operator $\widehat{R}(t)$ 
\begin{equation}
\frac{\partial \widehat{R}(t)}{\partial t}+\frac{i}{\hbar }\left[ \widehat{H},\widehat{R}(t)\right] =0.
\end{equation}
Here $\widehat{H}$ is the DQW many-band Hamiltonian we have described above.
If we project on the conduction band states we find the 2$\times $2 matrix
kinetic equation in the momentum representation%
\begin{eqnarray}
&&\frac{\partial \left[ \widehat{\rho }_{\mathbf{p}}(t)\right] _{jj^{\prime
}}}{\partial t}+\frac{i}{\hbar }\left[ \widehat{h},\widehat{\rho }_{\mathbf{p
}}(t)\right] _{jj^{\prime }}-\frac{i}{\hbar }\left\langle \left[
V_{c},a_{j^{\prime }s\mathbf{p}}^{\dagger }a_{js\mathbf{p}}\right]
\right\rangle  \nonumber \\
&=&\left[ \widehat{G}_{\mathbf{p}}(t)\right] _{jj^{\prime }}.
\end{eqnarray}
This equation describes the time evolution of the density matrix $\widehat{
\rho }_{\mathbf{p}}(t)$, where indexes $j,\ j^{\prime }$ refer to the levels 
$u$ (upper) and $l$ (lower) of the DQW, and $\mathbf{p}$ is the electronic
wave vector. \ In the expression (15) $\left[ \widehat{\rho }_{\mathbf{p}}(t)
\right] _{jj^{\prime }}=\left\langle a_{j^{\prime }s\mathbf{p}}^{\dagger
}a_{js\mathbf{p}}\right\rangle $, where $a_{j^{\prime }s\mathbf{p}}^{\dagger
}$ and $a_{js\mathbf{p}}$ are the creation and annihilation electron
operators, $s$ is the electronic spin, and $\left\langle \ \right\rangle $
means the statistical average. The operator $\widehat{h}=\left( \Delta
/2\right) \widehat{\sigma }_{z}+T\widehat{\sigma }_{x}$ is the matrix of the
one-electron DQW Hamiltonian without Coulomb interactions, and $\widehat{
\sigma }_{i}\ (i=x,y,z)$ are the standard Pauli matrices. For numerical and
analytical reasons it is convenient to remove the Coulomb interaction from
Hamiltonians (3) and (10) and to work it separately in the Hartree
representation. Thus, variational coefficients and normalization factors in $
\phi _{k}^{m}(\mathbf{\rho )}$ (13) will be slightly different. The Coulomb
interaction can be written as 
\begin{equation}
V_{c}=\frac{1}{2S}\dsum\limits_{ij}\dsum\limits_{ss^{\prime }}\dsum\limits_{
\mathbf{kk}^{\prime }\mathbf{q}}M_{ij}(q)a_{is\mathbf{k+q}}^{\dagger
}a_{js^{\prime }\mathbf{k}^{\prime }-\mathbf{q}}^{\dagger }a_{js^{\prime }
\mathbf{k}^{\prime }}a_{is\mathbf{k}},
\end{equation}
where $i,\ j$ refer both to electron and hole states to include
electron-electron, electron-hole and hole-hole Coulomb interactions. $S$ is
the normalization area and $q=\left\vert \mathbf{q}\right\vert $. The matrix
elements $M_{ij}(q)$ are defined as \cite{21}%
\begin{equation}
M_{ij}(q)=-\frac{2\pi e_{i}e_{j}}{\epsilon q}\diint dzdz^{\prime }\exp
\left( -q\left\vert z-z^{\prime }\right\vert \right) \left\vert \phi
_{i}(z)\right\vert ^{2}\left\vert \phi _{j}(z^{\prime })\right\vert ^{2},
\end{equation}
where $e_{i(j)}=-e$ for electrons $e_{i(j)}=e$\ for holes ($e$ is the
electron charge).

After substitution of (16) in (15) we reach 
\begin{equation}
\frac{\partial \widehat{\rho }_{\mathbf{p}}(t)}{\partial t}+\frac{i}{\hbar }
\left[ \widehat{h}+\widehat{Q}(t)+\widehat{X}_{\mathbf{p}}(t),\widehat{\rho }
_{\mathbf{p}}(t)\right] =\widehat{G}_{\mathbf{p}}(t),
\end{equation}
where $\widehat{Q}(t)$ is the direct Coulomb potential and $\widehat{X}_{%
\mathbf{p}}(t)$ is the exchange potential. If we define the projectors over
the $l$ and $u$ conduction states as $\widehat{P}_{l}=\left( 1+\widehat{%
\sigma }_{z}\right) $ and $\widehat{P}_{u}=\left( 1-\widehat{\sigma }%
_{z}\right) $, respectively, $n_{h}(t)$ is the hole concentration in the $h$%
-th hole subband, and $n_{l,u}(t)=\left( 2/S\right) \sum_{\mathbf{p}}Tr\left[
\widehat{P}_{l,u}(t)\widehat{\rho }(t)\right] $ is the electron
concentration in the upper ($u$) and lower ($l$) levels, thus matrices $%
\widehat{Q}(t)$ and $\widehat{X}_{\mathbf{p}}(t)$ can be written as
\begin{eqnarray}
\widehat{Q}(t) &=&\left[ M_{ll}(q)n_{l}(t)+M_{lu}(q)n_{u}(t)-%
\sum_{h}M_{lh}(q)n_{h}(t)\right] _{q\longrightarrow 0}\widehat{P}_{l}  \nonumber
\\
&&-\left[ M_{uu}(q)n_{u}(t)+M_{ul}(q)n_{l}(t)-\sum_{h}M_{lh}(q)n_{h}(t)%
\right] _{q\longrightarrow 0}\widehat{P}_{u},
\end{eqnarray}
\begin{equation}
\widehat{X}_{\mathbf{p}}(t)=\frac{1}{S}\sum\limits_{\mathbf{q}}\left\{ 
\begin{array}{c}
\begin{array}{c}
\left[ M_{ll}(q)-M_{lu}(q)\right] Tr\left( \widehat{P}_{l}\widehat{\rho }_{%
\mathbf{p-q}}(t)\right) \widehat{P}_{l} \\ 
+\left[ M_{uu}(q)-M_{lu}(q)\right] Tr\left( \widehat{P}_{u}\widehat{\rho }_{%
\mathbf{p-q}}(t)\right) \widehat{P}_{u}%
\end{array}
\\ 
+M_{lu}(q)\widehat{\rho }_{\mathbf{p-q}}(t)%
\end{array}
\right\} .
\end{equation}
After integrating (18) over momentum $\mathbf{p}$ one finds that the
non-diagonal part of the exchange potential , $M_{lu}(q)\widehat{\rho }_{%
\mathbf{p-q}}(t)$, does not contributes to the commutator in eq. (18). The
diagonal part of $\widehat{X}_{\mathbf{p}}(t)$, together with the Coulomb
potential $\widehat{Q}(t)$, provides us the renormalization of the levels
and, thus, the renormalization of the splitting $\Delta $ because these
local terms couple different subbands. \ It is important to point out that,
contrary to Ref. (3), in our case we are dealing with typical values of $q$
of the order of $10^{8}$ m$^{-1}$, which is the same order as the distance
between the centers of the wells $Z$. Thus, we cannot neglect the exchange
contribution.

Now, we define the concentration matrix as $\widehat{n}(t)=\left( 2/S\right)
\sum_{\mathbf{p}}\widehat{\rho }_{\mathbf{p}}(t)$ and integrate over $%
\mathbf{p}$ to obtain the isospin representation \cite{2,3}:%
\begin{equation}
\widehat{n}(t)=n_{0}(t)+\sum\limits_{i}n_{i}(t)\widehat{\sigma }%
_{i}(t)=n_{0}(t)+\mathbf{n}(t)\widehat{\mathbf{\sigma }}.
\end{equation}
After substituting in (18) and introducing the subindex $k$ to differ free
electron $\left( f\right) $, excitonic electron $\left( exc\right) $ and
trionic electron $\left( tr\right) $, the above Liouville (18) equation
leads to the Bloch-equation system for the evolution of the isospin density
vector $\mathbf{n}_{k}(t)=\left(
n_{k}^{x}(t),n_{k}^{y}(t),n_{k}^{z}(t)\right) $ and the scalar density $%
n_{k}^{0}(t):$ 
\begin{eqnarray}
\frac{\partial \mathbf{n}_{k}(t)}{\partial t}-\left[ \mathbf{L}_{k}(t)\times 
\mathbf{n}_{k}(t)\right] &=&\mathbf{G}_{k}(t)-\mathbf{\Lambda }_{k}(t), 
\nonumber \\
\frac{\partial n_{k}^{0}(t)}{\partial t} &=&N_{k}\delta (t)-S_{k}(t).
\end{eqnarray}
where $N_{k}\delta (t)-S_{k}(t)=(1/S)\sum_{\mathbf{p}}Tr\left( \widehat{G}_{%
\mathbf{p}k}(t)\right) $ and $\mathbf{G}_{k}(t)-\mathbf{\Lambda }%
_{k}(t)=(1/S)\sum\limits_{\mathbf{p}}Tr(\widehat{\mathbf{\sigma }}\widehat{G}%
_{\mathbf{p}k}(t)$. The nonlinear dynamic properties of the system are
described by $\mathbf{L}_{k}(t)=(2T_{k}/\hbar ,0,\Delta _{k}(t)/\hbar )$,
where $\Delta _{k}(t)$ is the renormalized time dependent level splitting
induced by the carrier density and $T_{k}$ is the tunneling matrix element. $%
\mathbf{G}_{k}(t)=(0,0,N_{k}\delta (t))$ is the generation term and $N_{k}$
denotes the initial density of electrons. In the present case, just at $t=0$%
, $N_{f}=N_{e}$ (injected electron density) and $N_{exc,tr}=0$. The terms $%
\mathbf{\Lambda }_{k}(t)$ and $S_{k}(t)$ include generation and annihilation
rates for excitons and trions. We will describe them below.$\ $ Vector
equations (22) split into three coupled Bloch system, corresponding to free
electrons, excitonic electrons and trionic electrons. The time dependent
energy splitting renormalization is determined (after an easy but tedious
algebra) by the many-body Coulomb interaction through the Hartre-Fock
approximation (19,20)
\begin{eqnarray}
\Delta _{exc(tr)}(t) &=&\Delta _{exc(tr)}+\frac{2\pi e^{2}}{\epsilon }\int
\int dzdz^{\prime }\left\vert z-z^{\prime }\right\vert  \nonumber \\
&&\times \left[ \left\vert \phi _{u,exc(tr)}(z^{\prime })\right\vert
^{2}-\left\vert \phi _{l,exc(tr)}(z^{\prime })\right\vert ^{2}\right]  \nonumber
\\
&&\times \left[ 
\begin{array}{c}
\left( \left\vert \phi _{u,exc(tr)}(z)\right\vert
^{2}n_{exc(tr)}^{u}(t)+\left\vert \phi _{l,exc(tr)}(z)\right\vert
^{2}n_{exc(tr)}^{l}(t)\right) \\ 
-\dsum\limits_{h}\left\vert \phi _{h}(z)\right\vert ^{2}n^{h}(t)%
\end{array}
\right]  \nonumber \\
&&-\alpha \frac{2\pi e^{2}}{\epsilon }\int \int dzdz^{\prime }\delta
(z-z^{\prime })\left[ \left\vert \phi _{l,exc(tr)}(z^{\prime })\right\vert
^{2}-\left\vert \phi _{u,exc(tr)}(z^{\prime })\right\vert ^{2}\right]  \nonumber
\\
&&\times \left[ 
\begin{array}{c}
\left\vert \phi _{l,exc(tr)}(z)\right\vert ^{2}n_{exc(tr)}^{l}(t) \\ 
-\left\vert \phi _{u,exc(tr)}(z)\right\vert ^{2}n_{exc(tr)}^{u}(t)
\end{array}
\right] ,
\end{eqnarray}
and
\begin{eqnarray}
\Delta _{f}(t) &=&\Delta _{f}+\frac{2\pi e^{2}}{\epsilon }\int \int
dzdz^{\prime }\left\vert z-z^{\prime }\right\vert  \nonumber \\
&&\times \left[ \left\vert \phi _{u,f}(z^{\prime })\right\vert
^{2}-\left\vert \phi _{l,f}(z^{\prime })\right\vert ^{2}\right]  \nonumber \\
&&\times \left[ \left\vert \phi _{u,f}(z)\right\vert
^{2}n_{f}^{u}(t)+\left\vert \phi _{l,f}(z)\right\vert ^{2}n_{f}^{l}(t)\right]
\nonumber \\
&&-\alpha \frac{2\pi e^{2}}{\epsilon }\int \int dzdz^{\prime }\delta
(z-z^{\prime })\left[ \left\vert \phi _{l,f}(z^{\prime })\right\vert
^{2}-\left\vert \phi _{u,f}(z^{\prime })\right\vert ^{2}\right]  \nonumber \\
&&\times \left[ 
\begin{array}{c}
\left\vert \phi _{l,f}(z)\right\vert ^{2}n_{f}^{l}(t) \\ 
-\left\vert \phi _{u,f}(z)\right\vert ^{2}n_{f}^{u}(t)
\end{array}
\right] ,
\end{eqnarray}
where $n_{k}^{u(l)}(t)$ and $n_{h}(t)$ \ are the electronic density in the
upper \ (lower) level and the hole density in the right-hand well,
respectively. \ The last terms in (23) and (24), proportional to the
coefficient $\alpha $, correspond to the exchange interaction contribution.
To calculate this contribution we need to introduce the carrier screening
since the use of the bare exchange potential may lead to divergent results
due to its singularity when $\mathbf{q\rightarrow 0}$. This means that the
energy renormalization caused by the exchange term may diverge. In an easy
approach, and considering parabolic subbands filled by degenerate electrons,
we use the statically screened Coulomb potential $V_{s}(\mathbf{q)=}\frac{%
4\pi e^{2}}{\epsilon (q+q_{s})}$, where $q_{s}=\frac{2me^{2}}{\epsilon \hbar
^{2}}$ is the screening wavevector. In this case, the screening exchange can
be written as $\sum\limits_{\mathbf{q}}V_{s}(\mathbf{q})\left[ \widehat{\rho 
}(\mathbf{p-q)}\right] _{jj}\mathbf{+}\frac{1}{2}\underset{r\rightarrow 0}{%
\lim }\left[ V_{s}(r)-V(r)\right] $. \ The last term corresponds to the
difference between the screened and unscreened Coulomb potential, in the
real space, at $r=0$. Integration over $\mathbf{p}$ and $\mathbf{q}$ of the
exchange commutator in (18) leads to corrections of the order of $11\%$ in
the level splitting. In the present work $\alpha \approx 12.1\mathring{A}$.
Exact calculations lead to three different values of the parameter $\alpha $%
, but differences (of about $10\%$) are not visible in results.

We will assume that only the first heavy hole level is occupied. It is worth
mentioning that excitonic and trionic electrons not only interact among
themselves and with holes but with free electrons as well. It is important
to remember that, for direct-created excitons in DQWs, holes and electrons
are generated in different wells. Equations (23, 24) differ from similar
expression in Ref. \cite{3} due to the exchange contribution.

If we denote by $N_{h\text{ }}$as the initial total hole density, then $%
n_{h}(t=0)=N_{h}$ and
\begin{equation}
n_{h}(t)=n_{exc}^{u}(t)+n_{exc}^{l}(t)+\frac{n_{tr}^{u}(t)+n_{tr}^{l}(t)}{2}.
\end{equation}. The factor $1/2$ in the last addendum appears because $n_{tr}^{j}$ denotes
the trionic electron density which is twice the hole density forming trions.
The scalar density $n_{k}^{0}(t)$ is the sheet electron density (the total
density of electrons in the DQW structure). Also, we can write%
\begin{eqnarray}
n_{k}^{0}(t) &=&n_{k}^{l}+n_{k}^{u}(t),  \nonumber \\
n_{k}^{z}(t) &=&n_{k}^{l}(t)-n_{k}^{u}(t).
\end{eqnarray}

Now we define $\tau $ as the relaxation time for the non-diagonal part of
the density matrix \cite{3,23}. This relaxation leads to the interlevel
redistribution of the electron density. We consider here the generation and
annihilation terms. Initially, free electrons are injected
quasi-instantaneously and lose their free condition to generate excitons and
trions. So we have two annihilation mechanisms expressed through the
generation rates $\Gamma _{exc}$ and $\Gamma _{tr}$, respectively. Moreover,
trions destruction gives rise to free electrons because the recombination of
one electron with the hole leaves an additional free electron. Thus, we have
an additional generation term for free electrons at a rate $1/\tau _{tr}$.
By writing densities $n_{f}^{i}(t)$ and $n_{exc,tr}^{i}(t)$ in units of $%
N_{e}$, equation (22) for free electrons becomes%
\begin{eqnarray}
\frac{d}{dt}n_{f}^{x}(t)+\frac{\Delta _{f}(t)}{\hbar }n_{f}^{y}(t)+\frac{1}{%
\tau }n_{f}^{x}(t) &=&0,  \nonumber \\
\frac{d}{dt}n_{f}^{y}(t)-\frac{\Delta _{f}(t)}{\hbar }n_{f}^{x}(t)+\frac{%
2T_{f}}{\hbar }n_{f}^{z}(t)+\frac{1}{\tau }n_{f}^{y}(t) &=&0,  \nonumber \\
\frac{d}{dt}n_{f}^{z}(t)-\frac{2T_{f}}{\hbar }n_{f}^{y}(t)-\frac{N_{f}}{N_{e}%
}\delta (t)-\frac{1}{\tau _{tr}}n_{tr}^{0}(t)+\Gamma _{tr}n_{m}(t)+\Gamma
_{exc}n_{h}(t) &=&0,  \nonumber \\
\frac{d}{dt}n_{f}^{0}(t)-\frac{N_{f}}{N_{e}}\delta (t)-\frac{1}{\tau _{tr}}%
n_{tr}^{0}(t)+\Gamma _{tr}n_{m}(t)+\Gamma _{exc}n_{h}(t) &=&0,
\end{eqnarray}%
where $n_{m}(t)=\min [n_{f}^{0}(t),n_{exc}^{0}(t)]$. In a similar way
excitonic electrons appear, after the instantaneous hole diffusion, at a
rate $\Gamma _{exc}$ and disappear due to trion generation and recombination
at rates $\Gamma _{tr}$ and $1/\tau _{exc}$, respectively. Thus, for
excitonic electrons, 
\begin{eqnarray}
\frac{d}{dt}n_{exc}^{x}(t)+\frac{\Delta _{exc}(t)}{\hbar }n_{exc}^{y}(t)+%
\frac{1}{\tau }n_{exc}^{x}(t) &=&0,  \nonumber \\
\frac{d}{dt}n_{exc}^{y}(t)-\frac{\Delta _{exc}(t)}{\hbar }n_{exc}^{x}(t)+%
\frac{2T_{exc}}{\hbar }n_{exc}^{z}(t)+\frac{1}{\tau }n_{exc}^{y}(t) &=&0, 
\nonumber \\
\frac{d}{dt}n_{exc}^{z}(t)-\frac{2T_{exc}}{\hbar }n_{exc}^{y}(t)-\Gamma
_{exc}n_{h}(t)+\frac{1}{\tau _{exc}}n_{exc}^{0}(t)+\Gamma _{tr}n_{m}(t) &=&0,
\nonumber \\
\frac{d}{dt}n_{exc}^{0}(t)-\Gamma _{exc}n_{h}(t)+\frac{1}{\tau _{exc}}%
n_{exc}^{0}(t)+\Gamma _{tr}n_{m}(t) &=&0,
\end{eqnarray}
and, for trionic electrons,
\begin{eqnarray}
\frac{d}{dt}n_{tr}^{x}(t)+\frac{\Delta _{tr}(t)}{\hbar }n_{tr}^{y}(t)+\frac{1%
}{\tau }n_{tr}^{x}(t) &=&0,  \nonumber \\
\frac{d}{dt}n_{tr}^{y}(t)-\frac{\Delta _{tr}(t)}{\hbar }n_{tr}^{x}(t)+\frac{%
2T_{tr}}{\hbar }n_{tr}^{z}(t)+\frac{1}{\tau }n_{tr}^{y}(t) &=&0,  \nonumber \\
\frac{d}{dt}n_{tr}^{z}(t)-\frac{2T_{tr}}{\hbar }n_{tr}^{y}(t)-\Gamma
_{tr}n_{m}(t)+\frac{1}{\tau _{tr}}n_{tr}^{0}(t) &=&0,  \nonumber \\
\frac{d}{dt}n_{tr}^{0}(t)-\Gamma _{tr}n_{m}(t)+\frac{1}{\tau _{tr}}%
n_{tr}^{0}(t) &=&0.
\end{eqnarray}%
In the above expressions $\Gamma _{tr}^{-1}$ is the mean time required for
the trion formation and $\tau _{exc}$ and $\tau _{tr}$ are the intrinsic
exciton and trion lifetimes. An extensive analysis about these times can be
found in the paper of Esser et al.\cite{12}. We have considered that the
process of exciton formation is much faster than the generation of trions
and the annihilation of both excitons and trions \cite{14}. Actually, $%
\Gamma _{tr}$ corresponds to the scattering rate of excitons with free
electrons, which gives rise to the trion formation. Experimentally, this
scattering formation rate is proportional to the free electron density, to
the in-plane exciton area (which contributes to the cross section of the
process), and to the kinetic energy of the two interacting species. In our
work we have used data for $GaAs-GaAlAs$ included in \cite{14} .

Equations (28), (29), and (30) allow us to get the total density values 
\begin{eqnarray}
n_{T}^{z}(t) &=&n_{f}^{z}(t)+n_{exc}^{z}(t)+n_{tr}^{z}(t),  \nonumber \\
n_{T}^{0}(t) &=&n_{f}^{0}(t)+n_{exc}^{0}(t)+n_{tr}^{0}(t),
\end{eqnarray}
where $n_{T}^{z}(t)$ is proportional to the total dipole moment.

\subsection{The current density}

Within the dipole approximation the current density can be written as 
\begin{equation}
j_{ll^{\prime }}(t)=\frac{e}{L}\sum\limits_{k}n_{k}^{z}(t)v_{\perp
ll^{\prime }}^{k},
\end{equation}
where \ $l,\ l^{\prime }$ refer to the coupled levels, $v_{\perp ll^{\prime
}}=-\frac{i\hbar }{m}\left\langle \phi _{l}(z)\right\vert \frac{\partial }{%
\partial z}\left\vert \phi _{l^{\prime }}(z)\right\rangle =v_{\perp lu}=$ $-%
\frac{i\hbar }{m}\left\langle \phi _{l}(z)\right\vert \frac{\partial }{%
\partial z}\left\vert \phi _{u}(z)\right\rangle $ is the intersubband
transverse velocity, and $L$ is the size of the structure. Since we have
supposed a left-hand barrier doping to locate the Fermi level just above the
two deepest conduction levels, only these two levels will be occupied and we
can neglect contributions from remote bands. Therefore, the whole problem is
now reduced to the additional calculation of the transverse velocity.

\section{Results and discussion}

We have used in calculations an asymmetric DQW (Fig.1) configured by a 110%
\AA\ $GaAs$ LQW and a 100\AA\ $GaAs$ RQW, separated by a 30\AA\ $%
Ga_{0.65}Al_{0.35}As$ barrier. Then we can guarantee the interaction of the
electron wave functions of neighboring wells (non-null overlap). For this
structure the distance between the centers of the wells is $Z_{c}=135$\AA .
We have also considered the permittivity as constant along the whole
structure, $\epsilon =12.9$. To analyze the contribution of trions to the
total dipole moment we perform numerically the coupled system (28-30)
through the Runge-Kutta method. We will restrict ourselves to cases in which
the density of injected electrons is higher than the density of diffused
holes, e.g., an excess of free electron density of about $%
10^{9}-10^{10}cm^{-2}$. Higher densities could prevent the trion formation 
\cite{11}.

Figures 3(a-c) represent the density evolution of free, excitonic and
trionic electrons [$n_{k}^{0}(t)$] for different $N_{h}/N_{e}$ ratios.
Initially, there are only free electrons. The density of these electrons
decays quickly, giving rise to the excitons. These excitons yield to trions
generation due to the scattering processes between excitons and free
electrons. The recombination of electron-hole pairs reduces the density of
electrons both in exciton and in trion. In the last case each extinct trion
leaves one electron free, which contributes to increase the stock of
remaining free electrons. That recombination leads to a long time limit of
the total density of electrons that equals the free electron excess, $%
n_{T}^{0}(t\rightarrow \infty )=n_{f}^{0}(t\rightarrow \infty
)=1-N_{h}/N_{e} $. Cases (a), (b), (c) corresponds to $N_{h}/N_{e}=0.8,\
0.5,\ 0.2$, respectively. At first sight, the formation of trions seems to
be more efficient when the initial hole concentration is half the initial
free electron concentration. In this case the trionic electron density can
even exceed the excitonic electron density.

\begin{figure}
\begin{center}
\includegraphics{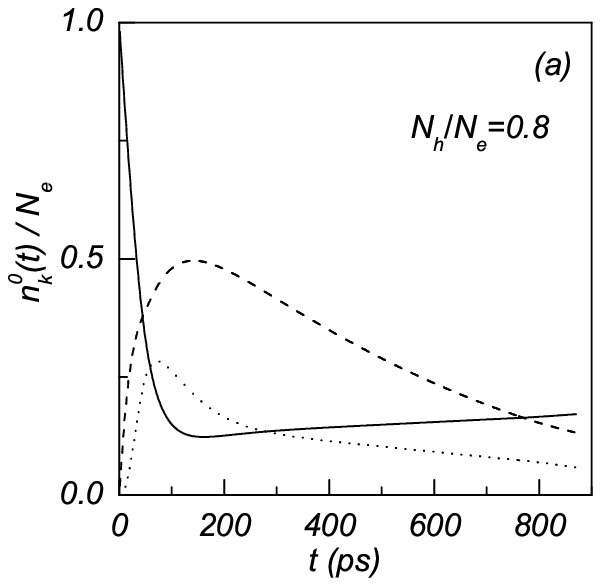} \includegraphics{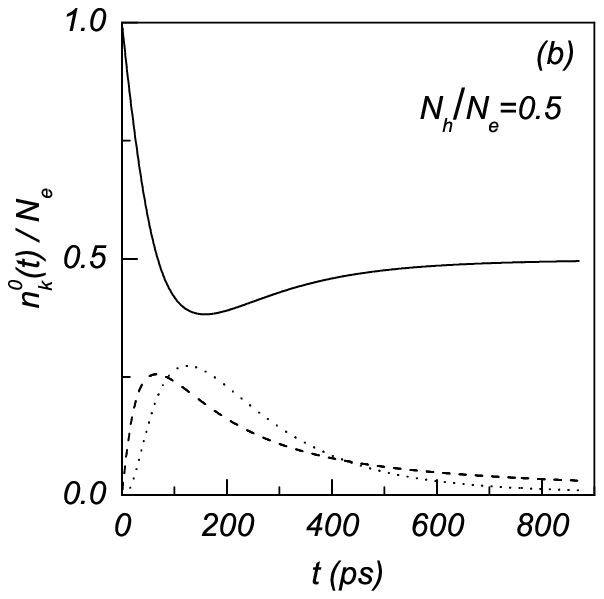}
\includegraphics{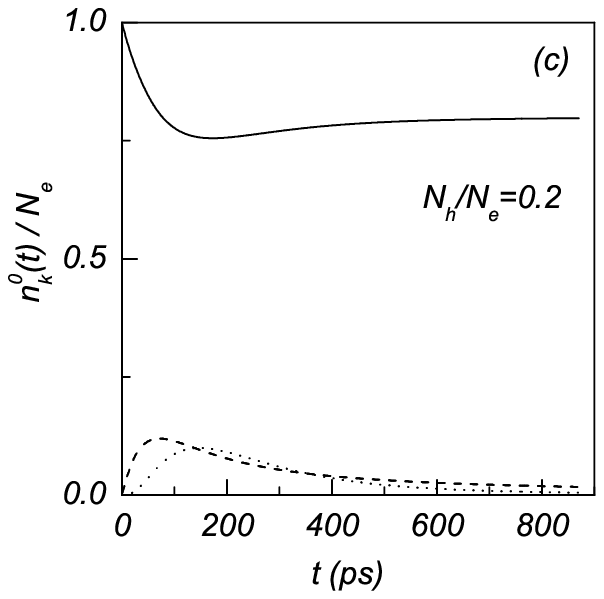}
\end{center}
\par
\addvspace{-1 cm}
\caption{Temporal evolution of the normalized electron density. (a) $%
N_{h}/N_{e}=0.8$, (b) $N_{h}/N_{e}=0.5$, and (c) $N_{h}/N_{e}=0.2$. Solid
line: free electrons, dashed line: excitons and dotted line: trions.}
\label{fig.3}
\end{figure}

Next we analyze the behavior of $n_{k}^{z}(t)$ to obtain $n_{T}^{z}(t)$,
which is proportional to the dipole moment. As stated in Eq. (26) $%
n_{k}^{z}(t)$ is the difference between the densities of the lower and the
upper energy levels. We represent $n_{k}^{z}(t)$ in Panels 4$\left(
a-c\right) $ for $N_{e}=10^{10}cm^{-2}$. Each figure corresponds to one of
the above mentioned $N_{h}/N_{e}$ ratios. As expected for low densities,
free electron density oscillates between lower and upper levels with an
exponential-like envelope profile not only due to the depopulation of the
corresponding levels but also to the interlevel redistribution. However,
Coulomb interaction forces the excitonic electrons to oscillate staying
basically in the upper level before the recombination and the interlevel
redistribution. Trionic electron tends to occupy mainly the lower level. On
the whole, the mentioned contributions give place to a modulation of the
free electron dipole moment which diminishes when $N_{h}/N_{e}$ does.

\begin{figure}
\begin{center}
\includegraphics{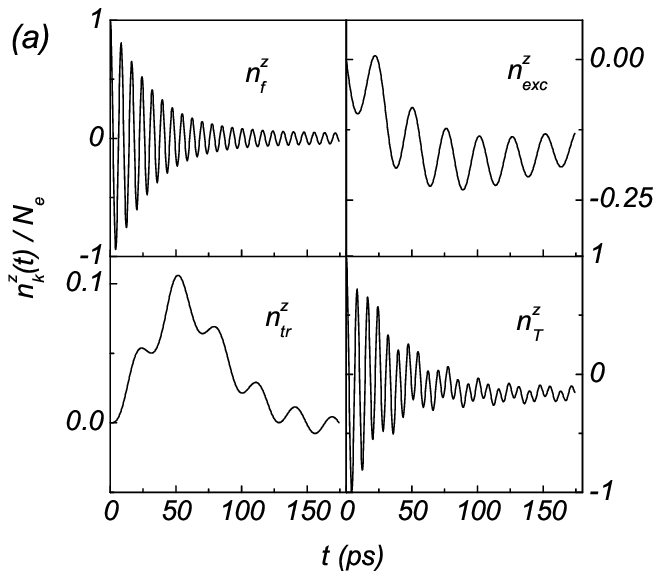}\includegraphics{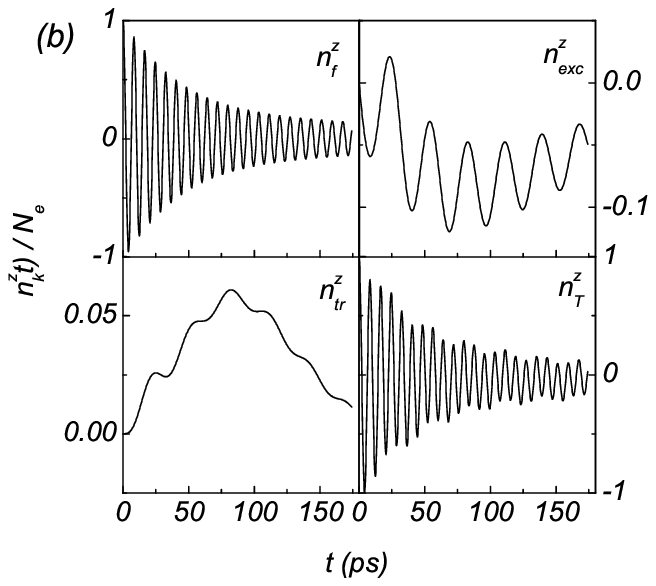} 
\includegraphics{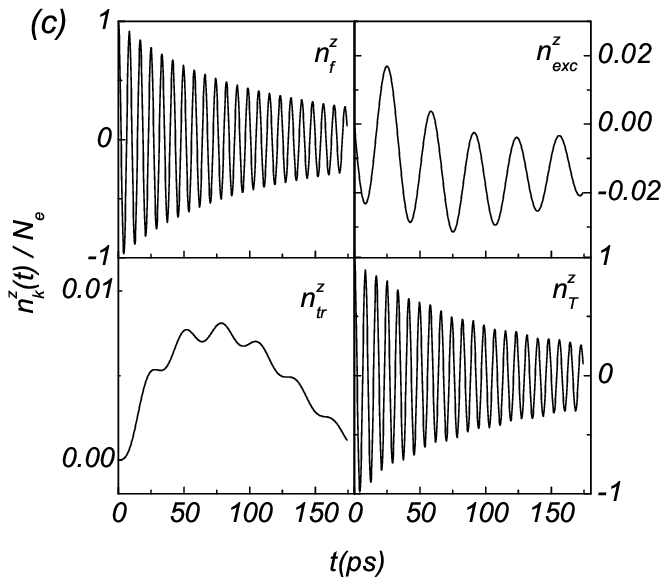}
\end{center}
\par
\addvspace{-1 cm}
\caption{Temporal evolution of the normalized dipole moment with $%
N_{e}=10^{10}cm^{-2}$. Upper panels: free and excitonic electrons. Lower
panels: trionic electrons and total dipole moment. (a) $N_{h}/N_{e}=0.8$,
(b) $N_{h}/N_{e}=0.5$, and (c) $N_{h}/N_{e}=0.2$.}
\label{fig.4}
\end{figure}

For higher densities the effect of the electron-electron Coulomb interaction
becomes more noticeable. Panels 5$\left( a-c\right) $ represent $%
n_{k}^{z}(t) $ for $N_{e}=5\times 10^{10}cm^{-2}$. Free electron density
behavior is absolutely different than the low density case and, initially,
most of the electrons stay in the lower level and only a few of them
oscillates between wells. At the same time excitonic electrons experience a
strong confinement in the upper level and trionic electrons do the same in
the lower level. The frequency of the three species of electrons shows an
increase that is more remarkable for the trion case. As in the low density
case, the effect over the free electron dipole moment due to excitons and
trions decreases when $N_{h}/N_{e}$ decreases. Another effect of the charge
density increase is to diminish the oscillation period breaking down the
harmonic response of the dipole moment. The anharmonic behavior can be
explained as a consequence of the space-charge potential created by the
spatial distribution of electrons and holes. This space-charge potential is
repulsive for holes and attractive for electrons. For densities higher than
the values used in this work ($N_{e}\gtrsim 10^{11}\ cm^{-2}$) the
Hartree-Fock potential becomes more important and a better approximation to
analyze the carrier distribution would be necessary. However, the
Hartree-Fock approximation used here for the many-body Coulomb potential
incorporates the main peculiarities of the particle interactions as function
of the carrier densities. Thus, electron-hole attraction dominates over
electron-electron and hole-hole repulsion at low densities. When the carrier
density increases the repulsive part of the Hartree-Fock potential increases
as well. Beyond certain initial density of free electrons, the repulsion
equals the attractive Coulomb interaction and the oscillation period
decreases for the three species. For higher densities the Coulomb
interaction of the second electron with the hole in the trion is canceled
resulting in the trion extinction, which leads to excitons and free
electrons. If the free electron density increases even more, the binding
energy of excitons tends to zero and excitons also disappear.

\begin{figure}
\begin{center}
\includegraphics{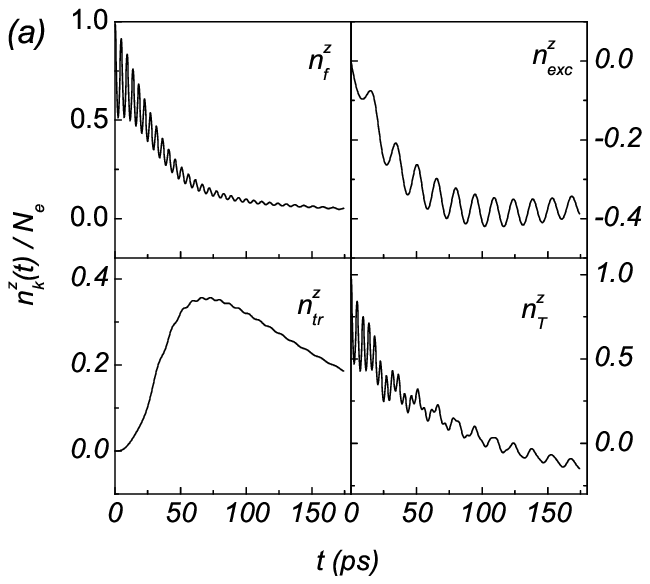} \includegraphics{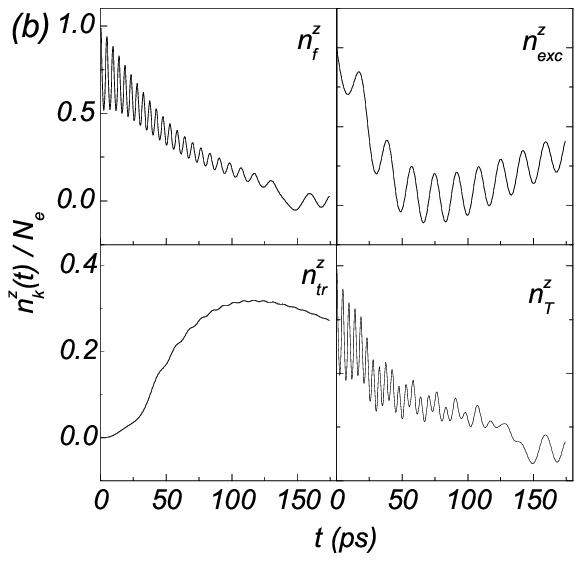}
\includegraphics{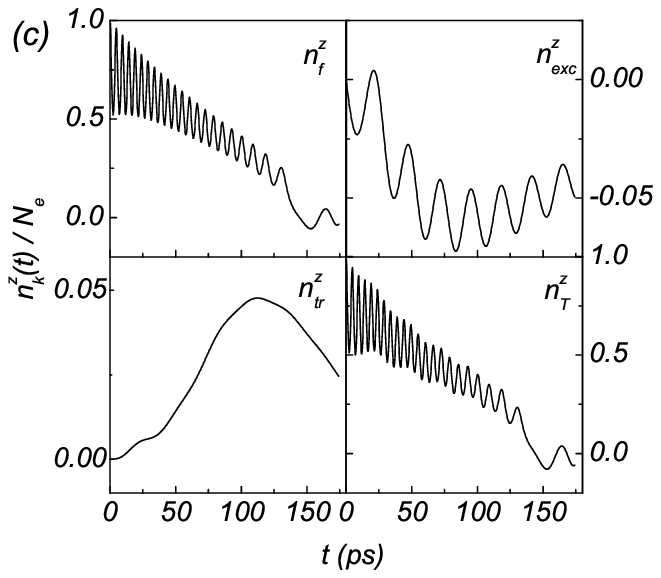}
\end{center}
\par
\addvspace{-1 cm}
\caption{Temporal evolution of the normalized dipole moment with $%
N_{e}=5\times 10^{10}cm^{-2}$. Upper panels: free and excitonic electrons.
Lower panels: trionic electrons and total dipole moment. (a) $%
N_{h}/N_{e}=0.8 $, (b) $N_{h}/N_{e}=0.5$, and (c) $N_{h}/N_{e}=0.2$.}
\label{fig.5}
\end{figure}

The shape of the current density is the same as the shape of the $%
n_{T}^{z}(t)$ density and we don't believe necessary to reproduce them here.
As an example we include current quantum beats in Fig. 6$\left( a\right) $
where upper panel corresponds to the current density for parameters of Fig. 4%
$(a)$, and lower panel to data of Fig. 5$(a)$. It should be notice that, in
any case, current density vanishes for long times due to relaxation as Fig. 6%
$\left( b\right) $ shows for the same parameters as in Fig. 6$\left(
a\right) $.

\begin{figure}
\begin{center}
\includegraphics{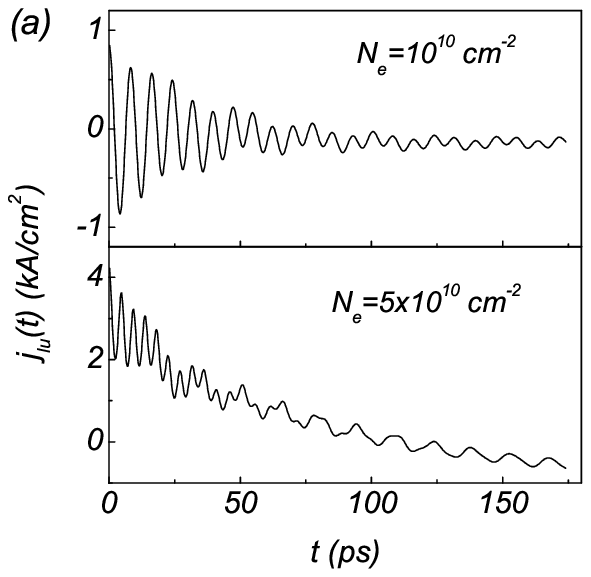}\includegraphics{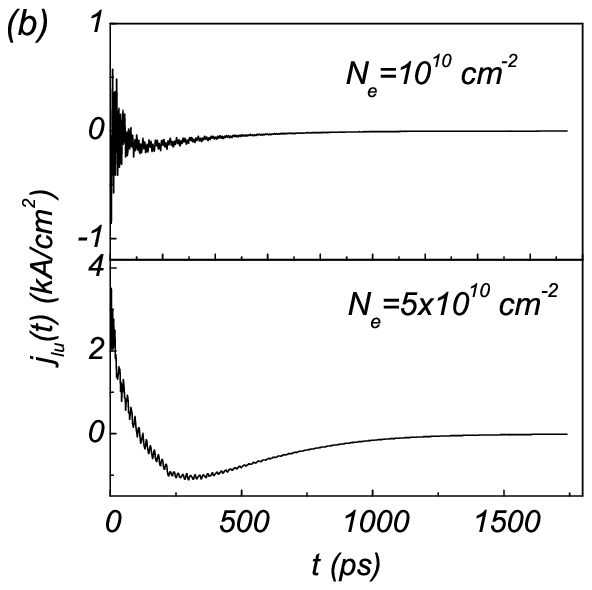}
\end{center}
\par
\addvspace{-1 cm}
\caption{Current density for $N_{h}/N_{e}=0.8$. Upper panel: $%
N_{e}=10^{10}cm^{-2}$. Lower panel: $N_{e}=5\times 10^{10}cm^{-2}.$ $\left(
a\right) $ quantum beats for short times. $\left( b\right) $ current
relaxation for long times.}
\label{fig.6a}
\end{figure}

Now, we will mention here the approximations used in the present method of
calculation. We assumed that the electron injection and hole diffusion are
processes much shorter than the coherent oscillation period and the exciton
generation time, which, in turn, is even shorter than trion generation and
recombination times. Another important point is the possible effect of the
applied external electric field on the trions. We have used field
intensities lower than $5 kV/cm$ to avoid the ionization or the possible
trion diffusion along the structure \cite{11b}.

In summary, we propose a new method for the generation of trions that avoids
the undesirable effects coming from the interaction of the laser
electromagnetic field with photoexcited carriers, an ever-present\ problem
in photoexcitation. The method is based on electron tunneling injection and
hole diffusion from remotely doped layers \cite{9}. We have also analyzed
the effect of these trions, together with the direct-created excitons, on
the time dependent dipole moment caused by the free electron density
oscillations. Such an effect is only relevant for an initial excess of the
free electron density with regard to the hole density. We have studied the
above mentioned effect through the three coupled Bloch system obtained from
the Liouville equation and taking into account five possible time dependent
processes. Our results show a wide variety of responses, from modulation and
quantum beats to anharmonic regime, caused by the different carrier
densities of the three types of electrons. These densities mainly affect the
dynamics of the level splitting energy of electrons and thus, the coupling
strength. The combined electron oscillations yield to a nonlinear coherent
electromagnetic radiation from the semiconductor structure. In other words,
the existence of trions could be detected by means of the current density,
which is straightforwardly translated into Terahertz radiation.

As far as we know only photoluminescence experiments for trions in coupled
DQW are available. Moreover, there are not any studies, neither experimental
nor theoretical, about the possible effect of trions on the dipolar
radiation emission from quantum wells. Therefore, we are forced to
extrapolate from well-known results for excitons. Since the standard way to
generate trions is ultra-fast photoexcitation, a detailed analysis of this
process deserves special attention together with the peculiarities of the
high-density case. We expect that the presented work will stimulate
experimental studies of the dynamics of direct-created excitons and trions.

\begin{acknowledgments}
This work has been supported in part by the Consejer\'{\i}a de Educaci\'{o}n,
Cultura y Deportes. Gobierno Aut\'{o}nomo de Canarias.
\end{acknowledgments}

\end{document}